\documentclass[12pt]{article}
\usepackage{bbm} 
\usepackage{amssymb}
\usepackage{graphicx}
\usepackage{amsmath}

\def\eeq{\end{eqnarray}}

\def\D{\mathcal{D}}

\def\im{\rm Im\, }
\def\=:{=\hspace{-.7em}\raisebox{1.1ex}{.}\hspace{.1em}\raisebox{-0.2ex}{.} }

\newcommand {\beq}{\begin{eqnarray}}
\newcommand {\eeqq}{\end{eqnarray}}
\newcommand {\non}{\nonumber\\}

\newcommand {\Tr}{{\rm Tr}\,}

\def\ket#1{|{#1}\rangle}

\def\brc{\langle}
\def\ckt{\rangle}
\def\de{\partial}

\makeatletter
\@addtoreset{equation}{section}

\renewcommand{\thefootnote}{\fnsymbol{footnote}}
\makeatother

\def\changed#1{{#1}}

\begin{document}

\pagestyle{empty}
\begin{flushright}
UT-Komaba/06-11, TIT/HEP-561, IFUP-TH/2006-26, RIKEN-TH-89, SISSA 76/2006/EP \\
{\tt hep-th/yymmnnn} \\
Nov, 2006 \\
\end{flushright}
\vspace{3mm}

\begin{center}
{\Large \bf
Non-Abelian duality  from  vortex moduli: \\
a dual model of  color-confinement }
\\[12mm]
\vspace{5mm}
{
 { \bf
Minoru~Eto}$^1$
\footnote{\it  e-mail address:
meto(at)hep1.c.u-tokyo.ac.jp
},
  { \bf
Luca~Ferretti}$^{2,3}$\footnote{\it  e-mail address:
l.ferretti(at)sissa.it
},
  { \bf
Kenichi~Konishi}$^{4,5}$\footnote{\it  e-mail address:
konishi(at)df.unipi.it
},
  { \bf
Giacomo~Marmorini}$^{6,5}$\footnote{\it  e-mail address:
g.marmorini(at)sns.it
},
  { \bf
Muneto~Nitta}$^7$
\footnote{\it  e-mail address:
nitta(at)phys-h.keio.ac.jp
},
  { \bf
 Keisuke~Ohashi}$^8$\footnote{\it  e-mail address:
keisuke(at)th.phys.titech.ac.jp
},
  { \bf
Walter~Vinci}$^{4,5}$\footnote{\it  e-mail address:
walter.vinci(at)pi.infn.it
}
  { \bf
 Naoto~Yokoi}$^{9}$\footnote{\it  e-mail address:
n.yokoi(at)riken.jp
},
}

\vskip 1.5em

{\small

$^1$ {\it
University of Tokyo,  Inst. of Physics,
Komaba 3-8-1, Meguro-ku
Tokyo 153, Japan
}
\\
$^2$ {\it SISSA,
via Beirut 2-4
I-34100 Trieste, Italy
}
\\
$^3$ {\it
INFN, Sezione di Trieste,
I-34012 Trieste (Padriciano),  Italy
}
\\
$^4$ {\it Department of Physics, University of Pisa \\
Largo Pontecorvo, 3,   Ed. C,  56127 Pisa, Italy
}
\\
$^5$ {\it INFN, Sezione di Pisa,
Largo Pontecorvo, 3, Ed. C, 56127 Pisa, Italy
}
\\
$^6$ {\it Scuola Normale Superiore,
Piazza dei Cavalieri, 7, 56126
 Pisa, Italy
}
\\
$^7$ {\it
Department of Physics, Keio University, Hiyoshi,
Yokohama, Kanagawa 223-8521, JAPAN
}
\\
$^8$ {\it Department of Physics, Tokyo Institute of
Technology,
Tokyo 152-8551, JAPAN}
\\
$^9$ {\it
Theoretical Physics Laboratory\\
The Institute of Physical and Chemical Research (RIKEN)\\
2-1 Hirosawa, Wako, Saitama 351-0198, JAPAN
}

} %
\vspace{10mm}
\newpage
{\bf Abstract}\\[5mm]
{\parbox{13cm}{\hspace{5mm}
{ It is argued that
the dual transformation  of non-Abelian monopoles occurring in a system with gauge symmetry breaking
$ G   \,\,\,\longrightarrow    \,\,\, H  $  is to be  defined by setting the low-energy  $H$ system in Higgs phase, so that the dual  system  is in confinement  phase.  The transformation  law of the  monopoles follows from that of  monopole-vortex mixed configurations in the system (with a large hierarchy of energy scales, $v_{1}\gg  v_{2}$)
\[
 G   \,\,\,{\stackrel {v_{1}} {\longrightarrow}} \,\,\, H  \,\,\,{\stackrel {v_{2}} {\longrightarrow}} \,\,\,
 {\mathbbm 1},
\]
under an unbroken, exact color-flavor diagonal symmetry $H_{C+F}\sim {\tilde H}$.
The transformation property  among  the  regular monopoles  characterized by $\pi_{2}(G/H)$,  follows from that  among the non-Abelian {\it vortices}  with flux  quantized according to $\pi_{1}(H)$,   via  the isomorphism  $\pi_{1}(G) \sim  \pi_{1}(H)/\pi_{2}(G/H)$.
Our  idea is  tested  against the concrete models  --
softly-broken ${\cal N}=2$ supersymmetric $SU(N), $ $SO(N)$ and $USp(2N)$  theories, with appropriate number of flavors.
The results
obtained in the semiclassical regime  (at $v_{1}\gg v_{2} \gg  \Lambda$)  of  these models  are consistent with those inferred from  the fully quantum-mechanical low-energy effective action  of the systems   (at $v_{1}, v_{2} \sim  \Lambda$).
}

}}
\end{center}
\vfill
\newpage
\setcounter{page}{1}
\setcounter{footnote}{0}
\renewcommand{\thefootnote}{\arabic{footnote}}

\pagestyle{plain}

\section{Introduction and discussion }\label{INTRO}

A system in which the gauge symmetry is spontaneously broken
\beq
  G   \,\,\,{\stackrel {\brc \phi_{1} \ckt    \ne 0} {\longrightarrow}} \,\,\, H   \label{this}
\eeqq
where $H$ is some non-Abelian subgroup of $G$,   possesses   a set of regular magnetic monopole
solutions  in the  semi-classical approximation,  which are natural generalizations  of the 't Hooft-Polyakov monopoles \cite{TH} found  in the system   $G=SO(3)$, $H=U(1)$.
A straightforward generalization of the Dirac's quantization condition leads to the GNOW (Goddard-Nuyts-Olive-E.Weinberg) conjecture, {\it i.e.}, that they form a multiplet of the group ${\tilde H}$,   dual of  $H$.  The group ${\tilde H}$ is generated  by  the dual  root vectors
\beq   \alpha^{*} = \frac{\alpha}{\alpha \cdot \alpha},
\eeqq
where $\alpha$ are the non-vanishing roots of $H$ \cite{GNO}-\cite{EW}.
There are however well-known difficulties in such an interpretation.  The first concerns the topological obstruction discussed
in \cite{CDyons}:  in the presence of the classical monopole background, it is not  possible  to define a globally  well-defined set of generators isomorphic to $H$.  As a consequence, no  ``colored dyons''  exist.  In the simplest example of a system with the symmetry breaking,
\beq  SU(3)  \,\,\,{\stackrel {\brc \phi_{1} \ckt    \ne 0} {\longrightarrow}} \,\,\, SU(2) \times U(1),
\label{simplebr}\eeqq
   this means that  no  monopoles exist which carry the quantum number, {\it e.g.},
    \beq  ({\underline 2},  1^{*}) \label{cannot} \eeqq
   where the asterisk indicates  the dual, magnetic  $U(1)$  charge.

The second  can be regarded as the  infinitesimal version of the same difficulty:   certain bosonic zero-modes around the monopole solution, corresponding to the $H$ gauge transformations,  are non-normalizable (behaving as $r^{-1/2}$ asymptotically).  Thus the standard procedure of \changed{semiclassical} quantization leading to the $H$  multiplet of the monopoles  does not work.    Some progress on the check of GNO duality along this  orthodox approach   has  been reported  nevertheless in \cite{DFHK} for ${\cal N}=4$  supersymmetric gauge theories, which however requires the consideration of particular multi-monopole systems  neutral with respect to the non-Abelian  group  (more precisely, non-Abelian part of)  $H$.

Both of these difficulties concern  the transformation properties of the  monopoles  under the subgroup  $H$, while the truly relevant question is how they transform under the dual group, ${\tilde H}$.  As field transformation groups, $H$ and ${\tilde H}$  are relatively non-local;    the latter  should look like a non-local transformation group  in the original, electric description.

Another related question concerns the multiplicity of the monopoles;   take again  the case  of the system with breaking pattern Eq.~(\ref{simplebr}).  One might argue that there is only one monopole, as  all the degenerate solutions  are related  by the unbroken {\it gauge}  group $H=SU(2)$.\footnote{This interpretation however encounters  the difficulties mentioned above. Also there are cases in which degenerate monopoles occur, which are not simply related  by the group $H$, see below.
}
Or one might say that there are two monopoles as, according to the semiclassical GNO classification, they are supposed to belong to a  doublet of the dual $SU(2)$ group.  Or, perhaps, one should conclude that there are infinitely many, continuously related solutions, as  the two solutions obtained by embedding the 't Hooft solutions in $(1,3)$ and $(2,3)$  subspaces, are clearly part of the continuous set of ({\it i.e.}, moduli of) solutions.  In short,  what is the multiplicity ($ \# $)  of the monopoles:
\beq    \#  \,\,  =  1, \quad 2,\quad  {\rm or} \quad   \infty \,\,  ?  \label{question}
\eeqq

Clearly the very concept of  the {\it dual gauge group } or  {\it dual gauge transformation}   must be better understood.  In attempting to gain such an improved insight on the nature of these objects, we are naturally  led to  several  general considerations.

The first is  the fact when  $H$ and $\tilde H $ groups are  non-Abelian  the  dynamics  of the system should enter the problem  in an essential  way. It should  not be surprising if  the understanding of  the concept of non-Abelian duality required a full   quantum mechanical treatment of the system.

For instance, the  non-Abelian $H$ interactions can become strongly-coupled at low energies and can break itself dynamically. This indeed occurs in pure
${\cal N}=2$ super Yang-Mills theories  ({\it i.e.}, theories without quark hypermultiplets), where the exact quantum mechanical
  result is known in terms of the Seiberg-Witten curves \cite{curves}.  Consider for instance, a pure ${\cal N}=2$,   $SU(N+1)$ gauge theory.  Even though partial breaking, {\it e.g.},  $SU(N+1) \to SU(N) \times U(1)$ looks perfectly possible semi-classically, in an appropriate region of classical degenerate vacua, no such vacua exist quantum mechanically.  In {\it all}   vacua the light monopoles are Abelian, the effective, magnetic  gauge group  being    $U(1)^{N}$.

  Generally speaking, the concept of a dual group multiplet  is well-defined only when ${\tilde H}$ interactions are weak (or, at worst, conformal).   This however means that one must study the original, electric theory in the regime of  strong coupling, which would usually  make the task of finding out
  what happens in the system at low energies exceedingly difficult.  Fortunately,    in ${\cal N}=2$ supersymmetric gauge theories,  the exact Seiberg-Witten  curves  describe the fully quantum mechanical consequences of the strong-interaction dynamics in terms of  weakly-coupled dual magnetic variables.   And this is  how we know that the non-Abelian monopoles do exist in fully quantum theories \cite{BK}:  in the so-called $r$-vacua of
softly broken ${\cal N}=2$ SQCD,  the light monopoles  interact  as a point-like particle in  a fundamental multiplet  ${\underline r}$ of the effective, dual $SU(r)$  gauge group.
  In the system of the type Eq.~(\ref{simplebr}) with appropriate number of quark multiplets  ($N_{f}  \ge 4$),  we know that  light magnetic monopoles carrying the non-Abelian quantum number  \beq  ({\underline {2^{*}}}, 1^{*}) \eeqq
  under the dual $SU(2) \times U(1)$
appear
  in the low-energy effective action  ({\it cfr.} Eq.~(\ref{cannot})).
  The distinction between $H$ and ${\tilde H}$ is crucial here.

  In general ${\cal N}=2$  SQCD   with  $N_{f}$ flavors,  light non-Abelian monopoles with $SU(r)$  dual gauge group appear  for   $r \le \frac{N_{f}}{2}$ only.  Such a limit clearly reflects the dynamical properties of the soliton monopoles under  renormalization group:  the effective low-energy  gauge group must be either infrared free or  conformal invariant, in order  for  the monopoles to emerge as  recognizable low-energy degrees of freedom \cite{APS}-\cite{BKM}.

 A closely related point concerns the phase of the system.   Even if  there is an ample evidence for the non-Abelian monopoles,  as explained above,
 we  might  still wish  to understand them in terms of something more familiar, such as semiclassical 't Hooft-Polyakov solitons.
 An analogous  question  can be (and should be)  asked about the Seiberg's ``dual quarks'' in ${\cal N}=1$ SQCD \cite{Seib}.
  Actually, the latter  can be interpreted  as the GNOW  monopoles becoming light due to the dynamics,   at least  in $SU(N)$  theories \cite{KMY}.   For $SO(N)$ or in $USp(2N)$ theories the relation between Seiberg duals and  GNOW  monopoles  are less clear \cite{KMY}.    For  instructive discussions on  the relation between Seiberg duals and  semiclassical monopoles  in a class of  ${\cal N}=1$ $SO(N)$  models with matter fields in vector and spinor representations,  see Strassler \cite{Strassler}.

Dynamics of the
system is thus  a crucial ingredient:  if the dual group were  in  Higgs phase, the multiplet  structure among the monopoles would get lost, generally.  Therefore one must study the dual  (${\tilde H}$)
system in  confinement phase.\footnote{
The non-Abelian monopoles in the  Coulomb phase suffer from the  difficulties already discussed.
}
{\it But then, according to the standard electromagnetic duality argument, one  must analyze the  electric  system   in Higgs phase. }  The monopoles will appear confined by the confining strings which are nothing but the vortices in the  $H$ system in  Higgs phase.

We are thus led to study   the system with a hierarchical symmetry breaking,
\beq
 G   \,\,\,{\stackrel {\brc \phi_{1} \ckt    \ne 0} {\longrightarrow}} \,\,\, H  \,\,\,{\stackrel {\brc \phi_{2} \ckt    \ne 0} {\longrightarrow}} \,\,\,
 \mathbbm {1},  \label{hierarchy}
\eeqq
where
\beq    |\brc \phi_{1} \ckt | \gg   |\brc \phi_{2} \ckt |,   \label{hierarchy2}
\eeqq
instead of the original system Eq.~(\ref{this}).
The smaller VEV breaks $H$ completely.  Also, in order for the  degeneracy among the monopoles not to be broken by the breaking at the scale $|\brc \phi_{2} \ckt |$, we assume
that some global color-flavor diagonal group
\beq   H_{C+F} \subset   H_{color} \otimes G_{F}   \label{symmetry}
\eeqq
remains unbroken.

It is hardly possible to emphasize the importance of the role of the massless  flavors too much.  This manifests in several different aspects.

\begin{description}
  \item[(i)] In order that $H$  must be non-asymptotically free, there must be sufficient number of massless flavors: otherwise, $H$ interactions would  become strong at low energies and $H$ group can break itself dynamically;
  \item[(ii)]  The physics of the $r$ vacua \cite{APS,CKM} indeed shows that the non-Abelian dual group $SU(r)$  appear only for $r \le \frac{N_{f}}{2}$.  This limit can be understood from the renormalization group:  in order for a non-trivial $r$ vacuum to exist, there must be at least  $2\, r $ massless  flavors in the fundamental theory;
   \item[(iii)]  Non-Abelian vortices \cite{HT,ABEKY}, which as we shall see are closely related to the concept of non-Abelian monopoles,
    require a flavor group.
The non-Abelian flux moduli arise as a result of an exact, unbroken color-flavor diagonal symmetry of the system,  broken by individual soliton vortex.

   \end{description}

The idea that the dual group transformations among the monopoles at the end of the vortices follow from those among the
vortices  (monopole-vortex flux matching, etc.), has been discussed  in several occasions, in particular in \cite{ABEK}.
The main aim of the present paper is to 
\changed{enforce this argument}, by showing that
the degenerate monopoles do indeed transform as a definite multiplet under a group transformation,  which is non-local in
 the original, electric variables, and involves flavor non-trivially,  even though this is not too obvious in the usual semiclassical
 treatment. The flavor dependence enters
through the infrared regulator.   The resulting, exact transformation group  is {\it defined}  to be
 the dual   group  of the monopoles.

\section{$SU(N+1)$  model with hierarchical symmetry breaking }

Our aim is to  show that all the difficulties about  the non-Abelian monopole moduli discussed in the Introduction  are eliminated by reducing the problem
 to that of the vortex moduli, related to the former by the topology and  symmetry argument.

\subsection{$U(N)$  model  with Fayet-Iliopoulos term \label{sec:UN} }

The model frequently considered  in the recent  literature in the  discussion of  various solitons  \cite{Tong}-\cite{SYSemi},  is a $U(N)$ theory with gauge fields $W_{\mu}$, an adjoint (complex) scalar $\phi$, and  $N_{f}=N$ scalar fields  in the fundamental representation of $SU(N)$,
with   the Lagrangian,
\beq
{\cal L} &=& \Tr \left[
- \frac{1}{2g^2} F_{\mu\nu}F^{\mu\nu}- \frac{2}{ g^2}  \D_\mu \,\phi^{\dagger} \,\D^\mu  \phi - \D_\mu \,H \,\D^\mu  H^\dagger
- \lambda \left( c\,{\bf 1}_N - H\,H^\dagger\right)^2\right] \nonumber  \\
&+&   \Tr \, [ \, (H^{\dagger} \phi  - M \,  H^{\dagger} )
(  \phi  \, H -  H \, M ) \,]  \label{widely}
\eeqq
where $F_{\mu\nu} = \partial_\mu W_\nu - \partial_\nu W_\mu + i \left[W_\mu,W_\nu\right]$
and $\D_\mu H  = \left(\partial_\mu + i\, W_\mu\right)\, H$,
and $H$  represents  the  fields in the fundamental representation of $SU(N)$,    written in  a color-flavor $N \times N$  matrix form,   $(H)_{\alpha}^{i} \equiv q_{\alpha}^{i}  $,  and $M$ is a $N\times N$ mass matrix.
Here, $g$ is the $U(N)_{\rm G}$ gauge coupling,  $\lambda$ is a
scalar coupling. For
\beq  \lambda = \frac{g^2}{4}\eeqq
  the system is BPS  saturated.
 For such a choice,  the model can be regarded as a  truncation
  \beq   (H)_{\alpha}^{i} \equiv q_{\alpha}^{i}, \qquad {\tilde q}^{\alpha}_{i}  \equiv 0   \eeqq
 of  the bosonic sector of an  ${\cal N}=2$ supersymmetric $U(N)$ gauge theory.  In the supersymmetric context the  parameter  $c$ is   the Fayet-Iliopoulos
parameter. In the following we set $c>0$  so that the system be in Higgs phase, and so as to allow stable vortex configurations.
For generic, unequal quark masses,
\beq    M =  diag \, (m_{1},m_{2},\ldots, m_{N}),   \label{unequalmcase}
\eeqq
the adjoint scalar VEV takes the form,
\beq   \brc \phi \ckt  =  M =   \left(\begin{array}{cccc}m_1 & 0 & 0 & 0 \\0 & m_2 & 0 & 0 \\0 & 0 & \ddots & 0 \\0 & 0 & 0 & m_N\end{array}\right),  \label{unequalmphi}  \eeqq
which breaks the gauge group to $U(1)^{N}$.
In  the equal mass case,
\beq    M =  diag \, (m,m,\ldots, m),   \label{equalmcase}
\eeqq
the adjoint and squark fields have the vacuum expectation value (VEV)
\beq   \brc \phi \ckt  =  m \, {\mathbf 1}_N,  \qquad \brc H  \ckt = \sqrt{ c} \,  \left(\begin{array}{ccc}1 & 0 & 0 \\0 & \ddots & 0  \\0 & 0 & 1   \end{array}\right).  \label{symmbr}
\eeqq
The squark VEV breaks the gauge symmetry completely, while leaving an unbroken $SU(N)_{C+F}$ color-flavor diagonal symmetry \changed{(remember that the flavor group acts on $H$ from the right
while the $U(N)_{\rm G}$  gauge symmetry acts on $H$ from the left)}.  The BPS vortex equations are
\beq
\left(\D_1+i\D_2\right) \, H = 0,\quad
F_{12} + \frac{g^2}{2} \left( c \,{\bf 1}_N - H\, H^\dagger\right) =0.
\eeqq
The matter  equation can be solved by use of the $N \times N$  moduli matrix $H_0(z)$
whose components are holomorphic functions of the  complex coordinate $z = x^1+ix^2$,
\cite{Isozumi:2004vg,Etou,Eto:2006pg}
\beq
H = S^{-1}(z,\bar z) \, H_0(z),\quad
W_1 + i\,W_2 = - 2\,i\,S^{-1}(z,\bar z) \, \bar\partial_z S(z,\bar z).
\eeqq
The gauge field equations then take the simple form  (``master equation'')  \cite{Isozumi:2004vg,Etou,Eto:2006pg}
\beq
\de_{z}\,(\Omega^{-1} \de_{\bar z} \, \Omega ) =  \frac{g^{2}}{4} \, ( c\, {\mathbf 1}_N -  \Omega^{-1} \, H_{0}\, H_{0}^{\dagger} ).
   \label{master} \eeqq
The  moduli matrix and $S$ are defined up to a  redefinition,
\beq   H_{0}(z)   \to  V(z) \, H_{0}(z), \qquad S(z, \bar{z})   \to  V(z) \, S(z, \bar{z}), \label{redef}
\eeqq
where  $V(z)$ is any non-singular  $N \times N$  matrix which is holomorphic in $z$.


\subsection {The Model  \label{sec:themodel} }

Actually the model we are interested here  is not exactly this model, but is a model which contains it  as a low-energy approximation.  We take as our model the  standard ${\cal N}=2$  SQCD  with  $N_{f}$ quark hypermultiplets, with a larger gauge symmetry, {\it e.g.},  $SU(N+1)$,   which is broken at a much larger mass scale as
\beq    SU(N+1)  \,\,\,{\stackrel  {v_{1}    \ne 0} {\longrightarrow}} \,\,\, \frac{ SU(N) \times U(1)}{{\mathbb Z}_{N}}.
\eeqq
The unbroken gauge symmetry is completely broken at a lower mass scale, as in Eq.~(\ref{symmbr}).

Clearly one can attempt  a similar embedding of the  model Eq.~(\ref{widely})  in a larger gauge group broken at some higher mass scale, in the context of a non-supersymmetric model,  even though in such a case the potential must be judiciously chosen and the dynamical  stability of the scenario would have to be carefully monitored.   Here we choose to study the softly broken ${\cal N}=2$ SQCD  for  concreteness,  and above all because  the dynamical properties of this model are well understood:  this will provide us with a non-trivial check of our results.  Another motivation is purely of convenience: it gives a definite potential with  desired properties.\footnote{
Recent developments \cite{DV,CDSW} allow us actually  to consider systems of this sort  within
 a much wider class of ${\cal N}=1$ supersymmetric models, whose infrared properties are very much under control.  We stick ourselves  to the standard ${\cal N}=2$ SQCD, however,  for concreteness.
}

The underlying theory is thus
\beq
{\cal L}=     \frac{1}{ 8 \pi} {\im} \, S_{cl} \left[\int d^4 \theta \,
\Phi^{\dagger} e^V \Phi +\int d^2 \theta\,\frac {1}{ 2} W W\right]
+ {\cal L}^{({\rm quarks})}  +  \int \, d^2 \theta \,\mu  \,\Tr  \Phi^2 + h.c. ;
\label{lagrangian}
\eeq
\beq {\cal L}^{({\rm quarks})}= \sum_i \, \left[ \int d^4 \theta \, \{ Q_i^{\dagger} e^V
Q_i + {\tilde Q_i}  e^{-V} {\tilde Q}_i^{\dagger} \} +  \int d^2 \theta
\, \{ \sqrt{2} {\tilde Q}_i \Phi Q_{i}    +      m_{i} \,    {\tilde Q}_i \, Q_{i}   \}+ h.c. \right]
\label{lagquark}
\eeq
where $m$ is the bare mass of the quarks and we have defined the complex coupling constant
\beq
S_{cl} \equiv  \frac{\theta_0}{\pi} +\frac {8 \pi i }{ g_0^2}.
\label{struc}
\eeqq
We also added the  parameter $\mu$, the mass of the adjoint chiral multiplet, which \changed{softly} breaks the supersymmetry to ${\cal N}=1$.
The bosonic sector of this model  is described, after elimination of  the auxiliary fields, by
\beq  {\cal L}=  \frac{ 1 }{4 g^2}  F_{\mu \nu}^2  +  \frac { 1}{ g^2}  |{\cal D}_{\mu} \Phi|^2 +
 \left|{\cal D}_{\mu}
Q\right|^2 + \left|{\cal D}_{\mu} \bar{\tilde{Q}}\right|^2 -    V_1 -  V_2,
\label{Lag}\eeqq
where
\beq    V_1   =          \frac { 1}{ 8 }  \, \sum_A  \left( t^A_{ij} \,[ \,  \frac { 1}{ g^2 }  (-2)  \, [\Phi^{\dagger},   \Phi]_{ji}  +  Q^{\dagger}_j   Q_i -  {\tilde Q}_j   {\tilde
Q}^{\dagger}_i \,] \right)^2;
\eeqq
\beq  V_2 &=&      g^2 |   \mu \, \Phi^A +
\sqrt 2   \, {\tilde Q} \, t^A  Q |^2    +     {\tilde Q } \,   [ m    + \sqrt2  \Phi   ] \,  [ m    + \sqrt2  \Phi   ]^{\dagger}  \, {\tilde Q}^{\dagger}
\nonumber  \\  & +  & Q^{\dagger} \, [ m    + \sqrt2  \Phi   ]^{\dagger}    \, [ m    + \sqrt2  \Phi   ]  \, Q.
\eeqq
In the construction of the approximate monopole and vortex  solutions  we shall consider only the VEVs and fluctuations around them which satisfy
\beq   [\Phi^{\dagger},   \Phi]=0, \qquad    Q_i   =  {\tilde Q}^{\dagger}_i,  \label{trunc1}
\eeqq
and hence  the $D$-term potential  $V_1$   can be set identically to zero throughout.

In order to keep the hierarchy of the gauge symmetry breaking scales, Eq.~(\ref{hierarchy2}), we
choose the masses such that
\beq     m_{1}=\ldots = m_{N_{f}}= m, \label{equalmass}  \eeqq
\beq    m   \gg  \mu    \gg   \Lambda.  \label{doublescale}
\eeqq
Although the theory described by the above Lagrangian has many degenerate vacua, we are interested in the vacuum where
(see \cite{CKM} for the detail)
\beq  \brc\Phi  \ckt =  - \frac{1}{\sqrt 2}  \, \left(\begin{array}{cccc} m & 0 & 0 & 0 \\0 & \ddots & \vdots  & \vdots \\0 & \hdots   & m & 0 \\0 & \hdots & 0 & - N \, m \end{array}\right);   \label{adjointvev}
\eeqq
\beq
Q=  {\tilde Q}^{\dagger}  = \left(\begin{array}{ccccc}  d & 0 & 0 &  0 & \hdots \\0 & \ddots  & 0 & \vdots & \hdots \\0 & 0 & d &  0 & \hdots \\0 & \hdots & 0 &  0 & \hdots\end{array}\right),  \qquad   d = \sqrt{ (N+1)\, \mu \, m}. \label{squarkvev} \eeqq
This is a particular case of the so-called $r$ vacuum, with $r=N$.  Although such a vacuum certainly exists classically, the existence of the  quantum $r=N$  vacuum in this theory requires  $N_{f} \ge  2\, N$,
which we shall assume.\footnote{
This might appear to be a rather tight condition as the original theory loses asymptotic freedom  for  $N_{f} \ge  2\, N + 2$.  This is not so.  An analogous discussion can be made by  considering the breaking
$SU(N) \to SU(r) \times U(1)^{N-r}$.   In this case  the condition for the quantum non-Abelian  vacuum is  $2\, N > N_{f} \ge  2\, r$, which  is a much looser condition.   Also,  although the corresponding $U(N)$  theory Eq.~(\ref{widely})  with such a number of flavor has  semilocal strings \cite{gibbons,Eto:2006pg,SYSemi}, these moduli are not directly related to the derivation of the dual gauge symmetry, which is our interest in this paper.  We shall come back to these questions elsewhere.
}

  To start with, ignore the smaller squark VEV, Eq.~(\ref{squarkvev}). As $\pi_{2}(G/H) \sim \pi_{1}(H) = \pi_{1}(SU(N)\times U(1)) ={\mathbb Z}$,
 the symmetry breaking  Eq.~(\ref{adjointvev})  gives rise to regular magnetic monopoles with mass of order of $O(\frac{v_{1}}{g})$,  whose continuous transformation property is our main concern here.  The semiclassical formulas for their mass and fluxes are well known  \cite{EW,ABEKM} and will not be repeated here.

\subsection {Low-energy approximation}

At scales much lower than  $v_{1} = m$ but still neglecting the smaller squark  VEV
$v_{2} =   d = \sqrt{ (N+1)\, \mu \, m} \ll  v_{1}$,    the theory reduces to an $SU(N)\times U(1)$  gauge theory with $N_{f}$ light quarks $q_{i}, {\tilde q}^{i}$ (the first $N$ components of the original quark multiplets $Q_{i}, {\tilde Q}^{i}$).  By integrating out the massive fields, the effective Lagrangian valid between the two mass scales has the form,
\beq  {\cal L} &=&  \frac{ 1}{ 4 g_N^2}  (F_{\mu \nu}^a)^2  + \frac{ 1}{ 4 g_1^2}  (F_{\mu \nu}^0)^2  +  \frac{ 1}{ g_N^2}  |{\cal D}_{\mu} \phi^a |^2 +
 \frac{ 1}{ g_1^2}  |{\cal D}_{\mu} \phi^0 |^2 +
 \left|{\cal D}_{\mu}
q \right|^2 + \left|{\cal D}_{\mu} \bar{\tilde{q}}\right|^2  \nonumber \\
&-&    g_1^2 \bigg|  -  \mu \, m  \sqrt{N(N+1)}  +
\dfrac{ {\tilde q} \,  q}{\sqrt{N(N+1)}}    \, \bigg|^2
 - g_N^2 |\sqrt 2   \, {\tilde q} \, t^a q \,  |^2  + \ldots
\label{leappr}    \eeqq
where $a=1,2,\ldots N^{2}-1$ labels the $SU(N)$ generators,  $t^a$;   the index $0$ refers to the $U(1)$ generator $t^0= \frac { 1 }{ \sqrt {2N(N+1)}} \, diag  (1, \ldots, 1, - N).$  We have taken into  account the fact that  the $SU(N)$ and $U(1)$   coupling constants ($g_N$ and   $g_1$) get renormalized differently towards the infrared.

 The adjoint scalars are fixed to its VEV,  Eq.~(\ref{adjointvev}), with small fluctuations around it,
\beq  \Phi =  \brc\Phi  \ckt  (1 +    \brc\Phi  \ckt^{-1} \, {\tilde  \Phi} )  , \qquad   |{\tilde  \Phi}| \ll m.
\label{small}\eeqq
In the consideration of the vortices of the low-energy theory,  they will be in fact replaced by the constant VEV.  The presence of the small terms Eq.~(\ref{small}), however, makes the low-energy vortices not strictly BPS  (and this will be  important in the consideration of their stability below).\footnote{In the terminology used in Davis et al. \cite{Davis}  in the discussion of the Abelian vortices in supersymmetric models, our model corresponds to an F model while the models of \cite{Tong,SY,Etou} correspond to a D model. In the approximation of replacing $\Phi$ with a constant,  the two models are equivalent: they are related by an $SU_{R}(2)$ transformation \cite{VainYung,ABE}.
}

The quark fields are replaced,  consistently with Eq.~(\ref{trunc1}),  as
\beq    {\tilde q} \equiv   q^{\dagger}, \qquad   q \to  \frac{1}{\sqrt{2}} \, q,
\eeqq
where the second replacement brings back the kinetic term to the standard form.

We further replace  the singlet coupling constant and the $U(1)$  gauge field as
\beq     e \equiv  \frac{g_{1}}{\sqrt{2 N(N+1)}};     \qquad   {\tilde A}_{\mu} \equiv   \frac{A_{\mu}}{\sqrt{2 N(N+1)}},  \qquad  {\tilde \phi}^{0}\equiv   \frac{\phi^{0}}{\sqrt{2 N(N+1)}}.
\eeqq
The net effect is
\beq  {\cal L} =  \frac{ 1}{ 4 g_N^2}  (F_{\mu \nu}^a)^2  + \frac{ 1}{ 4 e^2}  ({\tilde F}_{\mu \nu})^2  +
 \left|{\cal D}_{\mu}
q \right|^2
-    \frac{e^2}{2} \, |
 \, q^{\dagger} \,  q   -     c \, {\mathbf 1}  \, |^2 - \frac{1}{2} \, g_N^2 \,| \,
 \,  q^{\dagger} \, t^a q \,  |^2,
\eeqq
\beq  c=   2 N(N+1) \, \mu \, m.
\eeqq
Neglecting the small  terms left implicit, this  is  identical to the $U(N)$  model Eq.~(\ref{widely}), except for the fact that $e \ne g_{N} $ here.   The
transformation property of the vortices can be determined  from the moduli matrix, as was done in \cite{seven}.
Indeed, the system possesses BPS saturated vortices described by the linearized equations
\beq
\left(\D_1+i\D_2\right) \, q = 0,
\eeqq
\beq
F_{12}^{(0)} + \frac{e^2}{2} \left( c \,{\bf 1}_N - q\, q^\dagger \right) =0; \qquad F_{12}^{(a)} + \frac{g_{N}^2}{2}\, q_{i}^\dagger  \, t^{a}\, q_{i}  =0.
\eeqq
The matter equation can be solved exactly as in
\cite{Isozumi:2004vg,Etou,Eto:2006pg}  ($z = x^1+ix^2$) by setting
\beq
q  = S^{-1}(z,\bar z) \, H_0(z),\quad
A_1 + i\,A_2 = - 2\,i\,S^{-1}(z,\bar z) \, \bar\partial_z S(z,\bar z),
\eeqq
where $S$ is an  $N \times N$ invertible matrix  over whole of the $z$ plane, and  $H_{0}$ is  the  moduli matrix, holomorphic in $z$.

The gauge field equations take a slightly more complicated form than in the $U(N)$ model  Eq.~(\ref{widely}):
\beq  \de_{z}\,(\Omega^{-1}\de_{\bar z} \, \Omega ) =  -  \frac{g_{N}^{2}}{2}\, \Tr \, (\,t^{a} \, \Omega^{-1} \, q \, q^{\dagger} )\, t^{a} -
\frac{e^{2}}{4 N}\, \Tr \, (\,  \Omega^{-1} q \, q^{\dagger} - {\mathbf 1}), \qquad  \Omega =  S\, S^{\dagger}.
 \eeqq
The last equation reduces to the master equation  Eq.~(\ref{master}) in the  $U(N)$  limit, $g_{N}=e.$

The advantage of the moduli matrix formalism is that all the moduli parameters appear in the holomorphic, moduli matrix $H_{0}(z)$.   Especially, the transformation property of the vortices under the color-flavor diagonal group can be studied by studying the behavior of the moduli matrix.

\section {Topological stability, vortex-monopole complex   and  \\ confinement\label{sec:topology}}

The fact that there must be a continuous set of monopoles, which transform  under the color-flavor $G_{C+F}$ group,
follows from the following  exact  homotopy sequence
\beq    \cdots \to   \pi_{2}(G)   \to    \pi_{2}(G/H)   \to      \pi_{1}(H)  \stackrel{f}{\to}   \pi_{1} (G) \to \cdots,     \label{homotopy}
\eeqq
applied to our  systems    with a hierarchical symmetry breaking, Eq.~(\ref{hierarchy}),  with an exact unbroken symmetry,
Eq.~(\ref{symmetry}).   $\pi_{2}(G) =\mathbbm {1}$  for any Lie group,  and  $ \pi_{1} (G)  $ depends on the group considered.
Eq.~(\ref{homotopy}) was earlier used to obtain the relation between the regular, soliton monopoles (represented by
$\pi_{2}(G/H)$)  and the singular  Dirac monopoles, present if $ \pi_{1} (G)$ is non-trivial.      The isomorphism
\beq  \pi_{1}(G) \sim  \pi_{1}(H)/\pi_{2}(G/H)   \eeqq
  implied by Eq.~(\ref{homotopy})  shows that  among the magnetic monopole  configurations $A_{i}^{a}(x)$
 classified   according to $\pi_{1}(H)$ \cite{WuYang},   the regular monopoles  correspond to the
kernel of the map  $f: \pi_{1}(H)  \to   \pi_{1} (G)$ \cite{Coleman}.

When the homotopy sequence Eq.~(\ref{homotopy}) is applied to a  system with hierarchical breaking, in which $H$ is completely broken at low energies,
   \[ G   \,\,\,{\stackrel {v_{1}}  {\longrightarrow}} \,\,\, H  \,\,\,{\stackrel {v_{2}} {\longrightarrow}} \,\,\,
 \mathbbm {1},\]
   it allows  an interesting re-interpretation.   $\pi_{1}(H)$  classifies the quantized flux  of the vortices in the  low-energy  $H$  theory in Higgs phase.   Vice versa, the high-energy theory (in which  the small VEV is negligible) has 't Hooft-Polyakov   monopoles  quantized according to  $\pi_{2}(G/H)$.
However, there is something of a puzzle:   when the
small VEV's are taken into account,  which break  the ``unbroken'' gauge group completely,  these monopoles  must disappear somehow.    A related puzzle is that  the low-energy  vortices with    $\pi_{1}(H)$  flux,  would have to disappear in a theory where
$ \pi_{1} (G)$ is trivial.

\begin{figure}
\begin{center}
\includegraphics[width=3in]{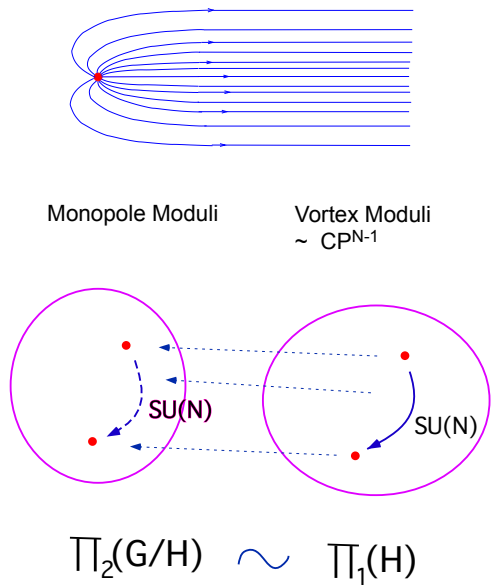}
\caption{\small The non-trivial vortex  moduli implies a corresponding  moduli of monopoles.}
\label{moduli}
\end{center}
\end{figure}

What happens is  that the massive monopoles are confined by the vortices and disappear from the spectrum;  on the other hand,  the vortices of the low-energy theory
end at the heavy monopoles  once the latter are taken into account, having mass large but not infinite (Fig.~\ref{moduli}).    The low-energy vortices  become unstable  also   through heavy
monopole pair productions  which break the vortices  in the middle (albeit with small, tunneling rates \cite{SYtun}),  which is really the same thing.
Note that,  even if  the effect of such  string breaking is neglected, a monopole-vortex-antimonopole configuration is not topologically stable anyway:  its energy would  become smaller if  the string becomes shorter (so such a composite, generally,    {\it will} get shorter and shorter and eventually disappear).

In the case   $G=SU(N+1)$,     $H=\frac{SU(N) \times U(1)}{\mathbb{Z}_{N}}$ we have a trivial
$\pi_{1}(G)$, so
   \beq   \pi_{2}\left(\frac{SU(N+1)}{U(N)}\right)
=  \pi_{2}({\bf C}P^{N})   \sim      \pi_{1}(U(N)) =  {\mathbb Z}:
    \eeqq
 each non-trivial element of $\pi_{1}(U(N)) $  is associated with a non-trivial element of $\pi_{2}(\frac{SU(N+1)}{U(N)}).$
Each vortex confines  a regular  monopole.   The monopole transformation properties follow from those of the vortices, as
will be more concretely studied in the next section.

In theories with a non-trivial $\pi_{1}(G)$ such as $SO(N)$,   the application of these ideas is slightly subtle:   these points will be discussed  in Section  \ref{sec:son}.

\begin{figure}
\begin{center}
\includegraphics[width=4in]{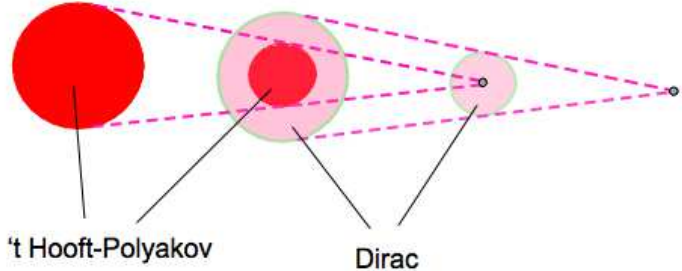}
\caption{\small A pictorial representation of the exact homotopy
sequence,  Eq.~(\ref{homotopy}),  with  the  leftmost figure
corresponding to  $\pi_{2}(G/H)$.} \label{sequence}
\end{center}
\end{figure}

In all cases,  as long as the group $H$ is completely broken at low energies and because $\pi_{2}(G)=\mathbbm {1}$ always,
none of the vortices (if $\pi_{1}(G)=\mathbbm {1}$)     and monopoles are truly stable, as static configurations.   They can be only approximately so, in an effective theory valid in respective regions  ($v_{1} \simeq \infty$  or   $v_{2}\simeq 0$).

 However, this does not mean that,  for instance, a  monopole-vortex-antimonopole composite configuration cannot be {\it dynamically} stabilized, or that they are not relevant as a physical configuration.  A rotation  can stabilize easily such a   configuration dynamically,   except that  it  will have a  small non-vanishing probability for decay  through a monopole-pair production,  if  such a decay is allowed kinematically.

  After all,  we believe that the real-world mesons are quark-string-antiquark bound states of this sort, the endpoints rotating almost with a speed of light!  An excited  meson  can and  indeed do decay through  quark pair productions into two lighter mesons (or sometimes to a baryon-antibaryon pair,  if allowed kinematically and by quantum numbers).  Only the lightest mesons are truly stable.  The same occurs with our monopole-vortex-antimonopole configurations.  The lightest such systems, after  the rotation modes are appropriately  quantized,   are   truly stable bound states of solitons, even though they might not be  stable as  static, semiclassical   configurations.


Our model is thus a reasonably faithful (dual)  model of the quark confinement in QCD.

A related point, more specific to the supersymmetric models we consider here as a concrete testing ground,  is the fact that
monopoles in the high-energy theory  and vortices in the low-energy theory,  are both BPS saturated.
It is crucial in our argument that  they  are both   BPS   only approximately;  {\it  they are almost BPS but not exactly}.\footnote{
The importance of almost BPS soliton configurations have also been emphasized by Strassler \cite{Strassler}.
}
They are unstable in the full theory.  But the fact that there exists a limit (of a large ratio of the mass scales, $\frac{v_{1}}{v_{2}}\to \infty$)   in which these solitons become exactly BPS and stable,
means that  the magnetic flux through the surface of a small sphere surrounding the monopole
and the vortex magnetic flux through a plane perpendicular to the vortex axis,  must match exactly.   These questions (the flux matching) have  been discussed extensively already  in \cite{ABEK}.

Our  argument, applied to the simplest case, $G= SO(3)$, and $H=U(1)$, is precisely the one adopted  by 't Hooft   in his pioneering paper \cite{TH} to argue that there must be  a regular monopole of charge two (with respect to the Dirac's minimum unit): as the vortex of winding number $k=2$  must be trivial in the full  theory ($\pi_{1}(SO(3))= \mathbb{Z}_{2}$),   such a vortex must end at a regular  monopole.
What is new here, as compared to the case discussed by 't Hooft \cite{TH}  is that now the unbroken group $H$ is non-Abelian and
that the low-energy vortices carry continuous, non-Abelian flux moduli.  The monopoles appearing as the endpoints of such vortices
must carry the same continuous moduli  (Fig.~\ref{moduli}).

The fact that the vortices of the low-energy theory are   BPS saturated (which allows us to analyze their moduli and transformation properties elegantly, as discussed in the next section), while in the full theory there are corrections which make them non BPS (and unstable), could cause some concern.  Actually,  the rigor of our argument is not
affected by those terms which can be treated as perturbation.  The attributes  characterized by integers such as   the transformation property of certain configurations as a multiplet of a  non-Abelian group which is an {\it exact symmetry group} of the full theory,  cannot receive renormalization. This is similar to the current algebra relations of Gell-Mann which are not renormalized.  Conserved vector current (CVC) of Feynman and Gell-Mann \cite{Feynman:1958ty} also hinges upon an analogous situation.\footnote{The absence of ``colored dyons''  \cite{CDyons} mentioned earlier  can also  be interpreted in this manner.}  The results obtained in the BPS limit  (in  the  limit  $v_{2}/v_{1} \to 0$)  are  thus  valid  at any  finite values of $v_{2}/v_{1}$.


\section {Dual gauge transformation among the monopoles}

The concepts such as the low-energy BPS vortices or the  high-energy BPS monopole solutions   are  thus only approximate:
their explicit forms are valid only in the lowest-order approximation, in the respective kinematical regions.   Nevertheless,  there is a property of the system   which  is exact   and does not depend on  any approximation: the full system has an exact, global $SU(N)_{C+F}$ symmetry,
which is neither broken by the interactions nor by both sets of  VEVs,  $v_{1}$ and  $v_{2}$.   This symmetry is    broken by individual soliton vortex,  endowing the latter  with  non-Abelian orientational moduli, analogous to the  translational zero-modes of a kink.
Note that the vortex breaks the color-flavor symmetry as
\beq   SU(N)_{C+F} \to  SU(N-1) \times U(1),
\eeqq
leading  to the moduli space of the minimum vortices  which is
\beq 
 {\cal M} \simeq  {\bf C}P^{N-1} = {SU(N) \over SU(N-1) \times U(1)}.
\eeqq
The fact that this moduli coincides with the moduli of the quantum states of an  $N$-state quantum mechanical system,  is a first hint that the monopoles appearing at the endpoint of a vortex, transform as a fundamental multiplet ${\underline N}$ of a group $SU(N)$.

The moduli space of the vortices is described by the moduli matrix  (we consider here the vortices of minimal winding,  $k=1$)
\beq   H_{0}(z)   \simeq  \left(\begin{array}{cccc}   1 & 0 & 0 & -a_1 \\   0 & \ddots & 0 & \vdots \\  0 & 0 & 1 & - a_{N-1} \\   0 & \ldots  & 0 & z\end{array}\right),
\label{minSUN}\eeqq
where the constants $a_{i}$, $i=1,2,\ldots, N-1$ are the coordinates of ${\bf C}P^{N-1}$.
Under $SU(N)_{C+F}$ transformation,  the squark fields transform as
\beq     q \to U^{-1} \, q \,U,
\eeqq
but as the moduli matrix is defined  {\it modulo}  holomorphic redefinition Eq.~(\ref{redef}), it is sufficient to consider
\beq     H_{0}(z)  \to    H_{0}(z)  \, U.
\eeqq
Now, for an infinitesimal $SU(N)$ transformation  acting on a matrix of the form Eq.~(\ref{minSUN}),  $U$ can be taken in the form,
\beq     U = {\bf 1} + X, \qquad X = \left(\begin{array}{cc}{\bf 0 }  & {\vec \xi} \\  -  ({\vec \xi})^{\dagger} & 0\end{array}\right),
\eeqq
where  ${\vec \xi}$ is a small $N-1$ component  constant vector.  Computing  $H_{0} \, X$ and  making a $V$ transformation from the left to bring back
$H_{0}$ to the original form,  we find
\beq     \delta a_{i} =   - \xi_{i}  -   a_{i} \, ({\vec \xi})^{\dagger}\cdot {\vec a},  \label{inhomo}
\eeqq
which shows that $a_{i}$'s indeed transform as the inhomogeneous coordinates of ${\bf C}P^{N-1}$.  In other words, the vortex represented by the moduli matrix  Eq.~(\ref{minSUN})  transforms as a fundamental multiplet of $SU(N)$.\footnote{
Note that,  if a ${\underline N}$  vector
${\vec c} $ transforms as ${\vec c}  \to  ({\bf 1} + X)\, {\vec c}  $,    the inhomogeneous coordinates   $a_{i} =  c_{i}/c_{N}$
transform as in Eq.~(\ref{inhomo}).
}

As an illustration consider the simplest case of $SU(2)$  theory.
In this case the  moduli matrix is simply \cite{Eto:2004rz}
\beq    H_{0}^{(1,0)} \simeq \left(\begin{array}{cc}z-z_0 & 0 \\  -b_{0}  & 1\end{array}\right); \qquad
 H_{0}^{(0,1)}  \simeq \left(\begin{array}{cc} 1  &  - a_{0}   \\ 0 & z-z_0\end{array}\right).
\label{minimum}\eeqq
with the transition function between the two  patches:
\beq b_{0}= \frac{1}{a_{0}}.  \label{simple}  \eeqq
The points on this ${\bf C}P^{1}$ represent all possible $k=1$ vortices.    Note that  points on the space of a quantum mechanical  two-state system,
 \beq  \ket {\Psi} = a_{1} \ket{\psi_{1}} + a_{2}\, \ket {\psi_{2}},  \label{twostate} \qquad     (a_{1}, a_{2}) \sim  \lambda \, (a_{1}, a_{2}), \quad \lambda \in {\bf C},
 \eeqq
 can be  put in one-to-one correspondence with  the inhomogeneous coordinate of a  ${\bf C}P^{1}$,
\beq      a_{0} =  \frac {a_{1}}{a_{2}}, \qquad    b_{0} =  \frac {a_{2}}{a_{1}}.  \label{correspondence}
\eeqq
In order to make this correspondence manifest, note that the minimal vortex Eq.~(\ref{minimum})  transforms under the $SU(2)_{C+F}$ transformation, as
\beq   H_{0} \to   V  \, H_{0}  \, U^{\dagger}, \qquad  U= \left(\begin{array}{cc}\alpha & \beta \\-\beta^* & \alpha^{*} \end{array}\right), \quad
|\alpha|^{2} + |\beta|^{2} =1,
\eeqq
where the factor $ U^{\dagger}$ from  the right represents a flavor transformation,
$V$ is a holomorphic matrix  which brings $H_{0}$ to the original triangular  form \cite{seven}.  The  action of this transformation on the moduli parameter, for instance, $a_{0}$,  can be found to be
\beq     a_{0} \to   \frac {\alpha \, a_{0} + \beta }{\alpha^{*} -  \beta^{*} \, a_{0}}.
\eeqq
But  this is precisely the way a doublet state  Eq.~(\ref{twostate})
  transforms under $SU(2)$,
\beq    \left(\begin{array}{c}a_1 \\a_2\end{array}\right)  \to    \left(\begin{array}{cc}\alpha & \beta \\-\beta^* & \alpha^{*} \end{array}\right) \, \left(\begin{array}{c}a_1 \\a_2\end{array}\right),
\eeqq

The fact that the  vortices (seen as solitons of the low-energy approximation)   transform as in the ${\underline N}$ representation of $SU(N)_{C+F}$, implies that
 there exist a set of monopoles  which transform accordingly, as   ${\underline N}$.  The existence of such a set follows from the exact $SU(N)_{C+F}$ symmetry of the theory, broken by the
individual   monopole-vortex configuration.  This answers  questions such as  Eq.~(\ref{question})  unambiguously.

{\it Note that in our derivation of continuous transformations of the monopoles,  the explicit, semiclassical form of the latter is not utilized. }

A subtle point is that  in the high-energy approximation, and to lowest order of such an approximation,   the semiclassical monopoles are just certain non-trivial field configurations
involving $\phi(x)$ and $A_{i}(x)$ fields, and  therefore apparently  transform under the color part of  $SU(N)_{C+F}$  only.  When the full
monopole-vortex configuration  $\phi(x), A_{i}(x), q(x) $  (Fig.~\ref{moduli}) are considered,  however,  only the combined color-flavor diagonal transformations keep  the energy of the configuration invariant.   In other words,  the monopole transformations must be regarded as  part of more complicated transformations involving  flavor,   when  higher order  effects  in $O(\frac{v_1}{v_2})$   are taken into account.\footnote{
Another  independent effect due to the massless flavors is that of Jackiw-Rebbi \cite{JR}: due to the normalizable zero-modes of the fermions, the semi-classical monopole is converted to some irreducible  multiplet of monopoles  in  the {\it flavor}  group $SU(N_{f})$. The ``clouds'' of the fermion fields surrounding the monopole have an  extension of  $O(\frac{1}{v_{1}})$, which is much smaller than the distance scales associated with the infrared effects discussed here and should be regarded as distinct effects.
}

And this means that the transformations are among physically distinct states, as the vortex moduli describe obviously physically distinct
vortices \cite{ABEKY}.



\subsection{ $SU(N)$ gauge symmetry breaking and Abelian monopole-vortex systems}

  Recently   there has been  considerable amount of research activity  \cite{HT},\cite{Tong}-\cite{SYSemi}, on systems  closely related to ours.  As the  terminology used and concepts  involved are often similar  but physically distinct, a confusion might possibly arise.

\begin{figure}
\begin{center}
\includegraphics[width=3in]{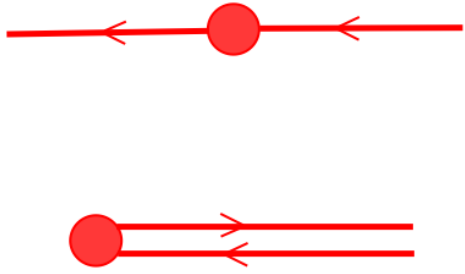}
\caption{Monopoles in  $U(N)$   systems with abelianization are confined by two Abelian  vortices. }
\label{two}
\end{center}
\end{figure}

As should be clear from what we said so far, it is crucial that the color-flavor diagonal symmetry $SU(N)$ remains  exactly conserved, for the emergence  of non-Abelian dual gauge group.  Consider, instead,  the cases in which  the gauge $U(N)$  (or $SU(N) \times U(1)$)  symmetry is broken to   Abelian subgroup  $U(1)^{N}$, either by small quark mass differences   ({\it cfr.}  Eq.~(\ref{unequalmphi})  and Eq.~(\ref{symmbr}))  or  dynamically,  as in the ${\cal N}=2$  models  with  $N_{f} <  2\, N$ \cite{HT2,SY}.     From the breaking of  various $SU(2)$ subgroups  to $U(1)$ there appear light 't Hooft-Polyakov  monopoles of mass  $O(\frac{\Delta m}{g})$ (in the case of an explicit breaking)  or   $O(\Lambda)$  (in the case of dynamical breaking).   As the  $U(1)^{N}$ gauge group is further broken by the squark VEVs,  the system develops ANO vortices.  The light magnetic monopoles,  carrying  magnetic  charges  of two  different  $U(1)$ factors,  look  confined  by the two vortices  (Fig.~\ref{two}).    These cases have been discussed extensively, within the context of $U(N)$  model of  Subsection \ref{sec:UN}, in \cite{HT},\cite{Tong}-\cite{Etou}. In Hanany et al.~\cite{Tong,HT2} and Shifman et al.~\cite{SY,GSY},  furthermore,  the dynamics of the fluctuation of  the orientational modes along the vortex, described as a two-dimensional  ${\bf C}P^{N-1}$ model,  is  studied.   It is shown that the kinks of the two-dimensional sigma model precisely correspond to these light monopoles, to be expected in the underlying $4D$ gauge theory.  In particular, it was noted that there is an elegant matching  between the dynamics of two-dimensional sigma model (describing the dynamics of the  vortex orientational modes in the Higgs phase of the $4D$ theory)   and the dynamics of the $4D$ gauge theory in the Coulomb phase \cite{Nick,Tong,HT2,SY}.

Note that this is also a reasonably  close (dual)  model of {\it what would occur}   in QCD if the color $SU(3)$  symmetry were to  dynamically break itself  to $U(1)^{2}$, {\it i.e.}, with generators  $Q^{1}= diag \, (1,-1,0)$,  $Q^{2}= diag \, (0,1,-1)$, respectively.   Confinement would be described in this case by the condensation of  magnetic monopoles carrying the Abelian charges $Q^{1}$, or $Q^{2}$, and the resulting ANO vortices will be of two types, $1$ and $2$  carrying the related fluxes.
   The quark  $q_{1}$  will be confined  by the vortex $1$,  the quark $q_{2}$   by the composite of the vortices   ${\bar 1}$ and $2$  (just as the light monopoles discussed above -- Fig.~\ref{two})  and  the quark $q_{3}$  by the vortex   ${\bar 2}$.

\subsection {Non-Abelian duality requires an exact flavor symmetry }

In the ${\cal N }=2$  supersymmetric QCD,
the presence of massless flavor and the exact color-flavor
diagonal symmetry is fundamental for the emergence of the dual (non-Abelian)  gauge transformations.  It is well  known in fact that the continuous non-Abelian vortex flux moduli - hence the non-Abelian vortex - disappear as soon as non-zero mass differences $m_{i}- m_{j}$ are introduced.\footnote{
Such an alignment of the vacuum with the bare mass parameters is characteristic of supersymmetric theories,  familiar also in the ${\cal N}=1$  SQCD
\cite{Amati}. In real QCD we do not expect such a strict alignment.
}
Also   in order for the $SU(N)_{C+F}$  color-flavor symmetry not to be destroyed by the gauge dynamics itself,  it is necessary to have   the number of flavors such that  $N_{f} \ge 2\, N$.
These points  have been emphasized already in the first paper on the subject \cite{ABEKY}.


It is illuminating  that the same phenomenon
  can  be seen  in   the fully quantum behavior of the theory of
Section \ref{sec:themodel},  in  another regime,
  \beq  \mu, m_{i}  \sim \Lambda  \eeqq
  ({\it cfr.}   Eq.~(\ref{doublescale})).    Indeed, this  model was analyzed thoroughly in this  regime in \cite{CKM}.   The so-called $r$ vacua
  with the low-energy effective $SU(r)\times U(1)^{N+1 -r}$  gauge symmetry emerges   in the equal mass limit $m_{i} \to m$  in which the global symmetry group $SU(N_{f}) \times U(1)$ of the  underlying theory become exact.  When the bare quark masses are almost equal but distinct,  the theory possesses a group of ${N_{f}\choose r}$  nearby vacua,  each of which  is an Abelian $U(1)^{N}$ theory,  with $N$   massless Abelian magnetic monopole pairs.    The jump from the $U(1)^{N}$ to $SU(r)\times U(1)^{N+1 -r}$  theory in the exact
 $SU(N_{f})$ limit might appear a discontinuous change of physics, but is not so.  What happens is that the range of validity of Abelian description in each Abelian vacuum, neglecting the light monopoles and gauge bosons (including massless particles of  the neighboring vacua, and other light particles which fill up a larger gauge multiplet in the limit the vacua coalesce), gradually tends to zero as the vacua collide.
The non-Abelian, enhanced gauge symmetry  of course  only emerges in the strictly degenerate  limit, in which the underlying theory has an exact $SU(N_{f})$ global symmetry.

\section {$SO(2N+1) \to SU(r) \times U(1)^{N-r-1}  \to \mathbbm {1}$ \label{sec:son} }


Let us now test  our ideas about duality transformations against another class of theories,
\beq  SO(2N+1)    \,\,\,{\stackrel {\brc \phi_{1} \ckt    \ne 0} {\longrightarrow}} \,\,\, SU(r) \times U(1)^{N-r+1}     \,\,\,{\stackrel {\brc \phi_{2} \ckt    \ne 0} {\longrightarrow}} \,\,\,\mathbbm {1}. \label{nonmaximal} \eeqq
One of the reasons why this case is interesting is that the semiclassical monopoles arising from the symmetry breaking
$ SO(2N+1)    \,\,\,{\stackrel {\brc \phi_{1} \ckt    \ne 0} {\longrightarrow}} \,\,\, U(N)$
 appear to  belong to  the  second-rank symmetric tensor representation  of $SU(N)$  \cite{DFHK,ABEKM}.
Another, related reason  is the fact that since $\pi_{1}(G)=\pi_{1}(SO(2N+1))= {\mathbb Z}_{2},$  the homotopy map Eq.~(\ref{homotopy}) is less trivial in this case.   Thirdly, according to the detailed analysis of the softly-broken ${\cal N}=2$  theories with $SO(N)$  gauge group \cite{CKKM}
 the quantum mechanical behavior of the monopoles is different for  $r=N$ and for $r < N$.  Non-Abelian  monopoles belonging to the fundamental representation of the dual $SU(r)$  group  appears only for $ r \le  N_{f}/2$, and because of the requirement of  asymptotic freedom of the original theory
 ($N_{f}   <   2N-1$),  this is possible only for $r < N.$    It is very encouraging that such a difference in the behavior of non-Abelian monopoles indeed follows, as we shall see,  from the way we define the dual group  though the transformation properties of  mixed monopole-vortex configurations  and homotopy map.

\subsection{ Maximal $SU$ factor;  $SO(5)\to U(2) \to \mathbbm {1}$}

Let us first  consider the case the $SU(N)$ factor has the maximum rank,
\[ SO(2N+1)    \,\,\,{\stackrel {\brc \phi_{1} \ckt    \ne 0} {\longrightarrow}} \,\,\, U(N). \]
To be concrete, let us consider the case of  an $SO(5)$ theory,  where a scalar VEV of the form
\beq  \brc \Phi \ckt  =   \left(\begin{array}{ccccc}0 & i  \,v & 0 & 0 & 0 \\- i  \,v & 0 & 0 & 0 & 0 \\0 & 0 & 0 & i  \,v  & 0 \\0 & 0 & - i  \,v  & 0 & 0 \\0 & 0 & 0 & 0 & 0\end{array}\right)
\label{SOvev}
\eeqq
breaking  the gauge group as  $SO(5) \to  H=SU(2) \times U(1) / {\mathbb Z}_{2}=U(2) $.   We assume that  at lower energies some other scalar VEVs break
$H$ completely, leaving however a color-flavor diagonal $SU(2)$ group unbroken.   This model arises   semiclassically  in softly broken ${\cal N}=2$  supersymmetric   $SO(5)$ gauge theory with  large, equal bare quark masses, $m$,  and with a small adjoint scalar mass $\mu$, with scalar VEVs given by $v=m/\sqrt{2}$ in Eq.~(\ref{SOvev}) and
\beq
Q=\tilde{Q}^{\dagger}=\sqrt{\frac{\mu m}{2}}\left(\begin{array}{cccc}1 & 0 & 0 & \cdots \\
i  & 0 & 0 & \cdots \\0 & 1 & 0 & \cdots \\0 & i & 0 & \cdots \\0 & 0 & 0 & \cdots
\end{array}\right).  \label{similar}
\eeqq
(See Appendix~\ref{a}, also  the Section 2 of \cite{CKKM},  for more details).

The $SO(4)\sim SU(2) \times SU(2)$  subgroup living on the upper-left corner is broken to $SU(2) \times U(1)$, giving rise to a single 't Hooft-Polyakov monopole.
On the other hand,  by embedding the  't Hooft-Polyakov monopole in the two $SO(3)$ subgroups  (in the $(125)$ and $(345)$  subspaces), one finds two more monopoles.  All three of them are degenerate.    Actually, E. Weinberg \cite{EW2}  has found a continuous set of degenerate monopole solutions  interpolating these, and noted that the transformations among them are not simply related to the unbroken $SU(2)$ group.\footnote{
This and similar  cases are  sometimes referred to as ``accidentally degenerate case'' in the literature.
}

From the  point of view of stability argument, Eq.~(\ref{homotopy}),  this  case is very similar to the case  considered by 't Hooft, as
$\pi_{1}(SO(5)) = {\mathbb Z}_{2}$:  a  singular  $\mathbb{Z}_{2}$ Dirac monopole can be introduced in the theory.  The minimal vortex of  the low-energy theory is truly stable in this case, as a minimal non-trivial element of $\pi_{1}(H)$ represents
also a non-trivial element of $\pi_{1}(G)$.   This can be seen as follows. A minimum element of $\pi_{1}(H) = \pi_{1}(U(2)) \sim {\mathbb Z}$  corresponds to  simultaneous
 rotations of angle $\pi$  in the $(12)$ and $(34)$ planes (which is a half circle of $U(1)$), which brings the origin to the ${\mathbb Z}_{2}$ element of
 $SU(2)$, $diag \, (-1,-1,-1,-1,1)$, followed by an $SU(2)$ transformation back to the origin, an angle $-\pi$ rotation in the $(12)$ plane and an angle $\pi$
 rotation around $(34)$ plane.  The net effect is a $2\pi$ rotation in the $(34)$ plane,  which is indeed a non-trivial element of $\pi_{1}(SO(5))={\mathbb Z}_{2}.$
Such a vortex would confine the singular Dirac monopole,  if introduced into the theory  (See Fig.~\ref{sequence}).

On the other hand, there are classes of vortices which appear to be stable in the low-energy approximation, but are not  so in the full theory.   In fact  non-minimal  $k=2$ elements of  $\pi_{1}(H) = \pi_{1}(SU(2) \times U(1) / {\mathbb Z}_{2}) \sim {\mathbb Z}$  are actually trivial in the full theory. This means that the $k=2$ vortices must end at a regular monopole.  Vice versa, as  $\pi_{2}(SO(5))=\mathbbm {1}$,  the regular 't Hooft Polyakov monopoles of high-energy theory  must be confined by these   non-minimal vortices and disappear from the spectrum.

The transformation property of $k=2$ vortices has been  studied recently in \cite{Hashimoto:2005hi,Auzzi:2005gr},  and in particular, in \cite{seven}. It turns out that the moduli space of the $k=2$  vortices is a  ${\bf C}P^{2}$ with a conic singularity.  It was shown that  the generic  $k=2$  vortices transform under  the $SU(2)_{C+F}$ group as a {\it  triplet}.   At a particular point of the moduli - an orbifold singularity - the vortex is Abelian:  it is a {\it singlet} of $SU(2)_{C+F}$. \footnote { In another complex codimension-one subspace,   they appear to  transform as  a {\it doublet}. However   quantum states of  {\it any} triplet  of $SU(2)$  contains such an orbit.  The state of maximum $S_{z}$, $\ket {1,1}$,  transforms under $SU(2)$  as an $SO(3)$ vector,  staying on a subspace $S^{2} \sim CP^{1}  \subset CP^{2}.$
  }

As the full theory has an exact, unbroken $SU(2)_{C+F}$ symmetry,
 it follows from the homotopy-group  argument of Section \ref{sec:topology} that {\it  the monopoles in the high-energy   $SO(5) \to  U(2)$  theory  have components transforming as a {\it triplet}    and a {\it singlet} of   $SU(2)_{C+F}$}.


 Note that it is not  easy to see   this result  -- and is somewhat misleading to attempt to do so   --    based solely on the semi-classical construction of the monopoles or on the zero-mode analysis around such solutions,  where the unbroken color-flavor symmetry is not appropriately taken into account.   Generically, the ``unbroken'' color  $SU(2)$ group suffers from the topological obstruction
 \cite{CDyons}   (or perturbatively,   from the  pathology of non-normalizable  gauge zero-modes \cite{CDyons,DFHK}), as we  noted already.

\begin{figure}
\begin{center}
\includegraphics[width=3in]{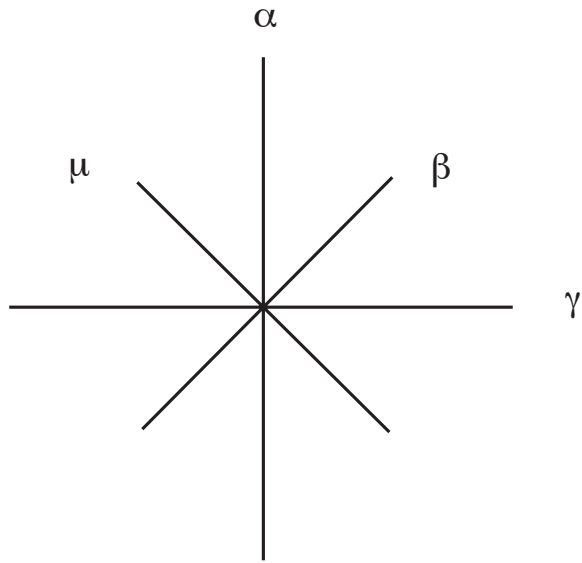}
\caption{Non-zero root vectors of $SO(5)$}
\label{roots}
\end{center}
\end{figure}

\begin{figure}
\begin{center}
\includegraphics[width=4in]{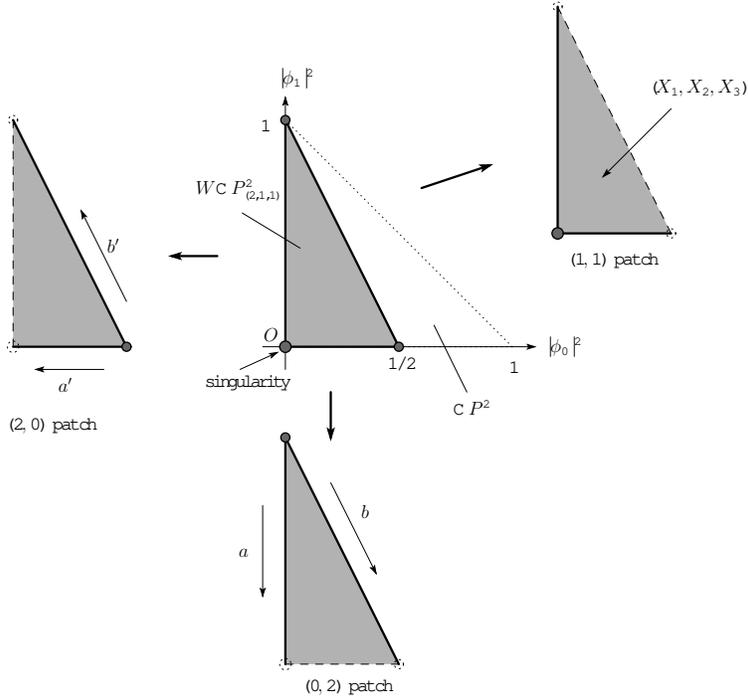}
\caption{Moduli space of $k=2$  vortices of $U(2)$  theory.  See \cite{seven} for more details.}
\label{Eto}
\end{center}
\end{figure}

Nevertheless, there are indications  that the findings by E. Weinberg \cite{EW2} are consistent with the properties of the $k=2$ vortices. In the standard way to embed $SU(2)$  subgroups through the Cartan decomposition (we follow here the notation of \cite{EW2}),
\beq    t_{1}(\nu) =  \frac{1}{(2\, \nu^{2})^{-1/2}}  \, (E_{\nu} + E_{-\nu}); \quad
 t_{2}(\nu) =   \frac{-i}{(2\, \nu^{2})^{-1/2}}  \, (E_{\nu} -  E_{-\nu}); \quad  t_{3}=   (\nu^{2})^{-1}\, \nu_{j}\, T_{j},
\eeqq
where $\nu$ denotes the non-vanishing root vectors of $SO(5)$  (Fig.~\ref{roots}),  the unbroken $SU(2)$ group is generated by $\gamma$.    The monopole associated with the root vector $\beta$ and the (equivalent) one given by $\mu$
naturally form a doublet of the ``unbroken''  $SU(2)$, while the monopole with the $\alpha$   charges
is a singlet.   The continuous set of monopoles interpolating among these monopoles found by Weinberg are analogous to the continuous set of vortices we found, which form the points of the ${\bf C}P^{2}$, which transform as a triplet.
(See the Fig.~\ref{Eto}  taken from \cite{seven}).

An even more concrete hint of consistency comes from the structure of the moduli space of the monopoles.  The moduli metric found  in \cite{EW2}  is
\beq  ds^{2} =   M\, d \, {\bf x}^{2} +   \frac{16 \pi^{2}}{M}  d\, \chi^{2}  + k\,
\left[ \frac{d\, b^{2}}{b} + b \, ( d\alpha^{2}+  \sin^{2}\alpha \, d \beta^{2} +  ( d \, \gamma + \cos \alpha \, d\, \beta)^{2}  )
\right].
\label{ewein}   \eeqq
{By performing a simple change of coordinate, $B\equiv 2 \sqrt{b}$, it becomes evident that    the moduli space  has  the structure}
\beq      {\bf C}^{2}/{\mathbb Z}_{2},  \label{metricEW}
\eeqq
 apart from the irrelevant factor ${\bf R}^{3}$ (the position of the monopole) and $S^{1}$  ($U(1)$ phase).\footnote{
The monopole modulus due  to the unbroken $U(1)\subset U(2)$ is not present in the full system, where the gauge group is completely broken.
}
Eq.~(\ref{metricEW})  coincides with the  moduli space of  the $k=2$  {co-axial {\it vortices}}, seen in the central $(1,1)$ patch \cite{seven}.



These considerations strengthen  our conclusion that the continuous set of monopoles found in \cite{EW2}  belongs to a singlet and a triplet  representations of the dual $SU(2)$  group.    Although the detailed properties of the moduli spaces for monopoles and vortices are different\footnote{
The first is known to be hyper-K\"ahler and the second K\"ahler  -- indeed ${\bf C}P^2/\mathbb{Z}_2$ does not admit hyper-K\"ahler structure.},  this could be related to the fact that
one should ultimately consider a smooth monopole-vortex  mixed configurations in the full theory,   not each of them separately.  Also, related to this point, there remains the fact that the dual group which is  exact and under which  monopoles transform,  is {\it not}  the original $SU(2)$ subgroup but involves the flavor group essentially.

Note that our conclusion is based on the exact symmetry, and should be reliable.   However,  the degeneracy among all  the vortices (or the monopoles) lying in  the entire moduli space ${\bf C}P^{2}/\mathbb{Z}_{2}$  found in the BPS limits,  is an artifact of the lowest-order approximation.
Only the degeneracy among the vortices (or among  the monopoles) belonging to the same multiplet  is expected to survive quantum mechanically.  ${\underline 1} $   and ${\underline 3}$  vortex tensions (monopole masses)  will split.   Which of the multiplets   (${\underline 1}$ or  ${\underline 3}$)  will remain stable, after quantum corrections  are taken into account,   is a question just lying beyond the power of semiclassical considerations.

  In  the context of  asymptotically-free ${\cal N}=2$  supersymmetric models,  there are no  indications that the triplet monopoles of   $SO(5) \to U(2) $  theory survive quantum mechanically.   This result can be  actually  understood by a simple renormalization-group argument:

\begin{itemize}
  \item
     In  a $SO(2N+1)$ theories   with ${\cal N}=2, 1$ supersymmetries,  the condition for the original theory to be  asymptotic-free
 ($N_{f}$    less   than  $2N-1$, $ \frac{3 (2N-2)}{2}$, respectively)\footnote{The counting is made for the appropriate supersymmetry multiplets,   $N_{f}$ hypermultiplets for ${\cal N}=2$;  $N_{f}$ chiral multiplets for ${\cal N}=1$  supersymmetric $SO(N)$ theory. }   is not compatible with the
 low-energy  $SU(N)$  theory being non-asymptotic-free  ($N_{f} \ge  2 N$ and   $N_{f} \ge   3 N$,  respectively.)
\end{itemize}

The problem would not arise if the rank of the unbroken $SU(r)$  were smaller.
That such a ``sign-flip'' of the beta function is a necessary condition for the emergence of low-energy non-Abelian monopoles has been pointed out some time ago by one of the authors \cite{KenNag},  even though  the validity of such an argument for non-supersymmetric theories is perhaps not obvious.

 If the condition of asymptotic freedom of the ultraviolet theory is dropped, then there are no such constraints, and  it makes sense to consider  symmetry breaking patterns  such as $SO(2N+1) \to U(N) $.   Our conclusion that the    monopoles   of  $SO(5) \to U(2)$  system  transform as a triplet or a singlet would apply under such conditions.  Analogously, we expect the monopoles in the system
 $SO(2N+1) \to U(N) $ to transform as a second-rank symmetric or antisymmetric representation.


%
%

\subsection{$SO(2N+1) \to SU(r) \times U(1)^{N-r-1}  \to \mathbbm {1}$        ($r <N$)}

Consider now the cases  in which the unbroken $SU(r)$ factor has a smaller rank,   $SO(2N+1) \to SU(r) \times U(1)^{N-r+1}  \to \mathbbm {1}$,  where    $r < N$.
For concreteness,  let us discuss the case of an $SO(7)$ theory,
\beq  SO(7)    \,\,\,{\stackrel {\brc \phi_{1} \ckt    \ne 0} {\longrightarrow}} \,\,\, U(2) \times U(1)    \,\,\,{\stackrel {\brc \phi_{2} \ckt    \ne 0} {\longrightarrow}} \,\,\,\mathbbm {1}. \label{nonmaxbis} \eeqq
As we are interested in a concrete dynamical realization of this,  we consider the softly broken
${\cal N}=2$  theory, with $N_{f}=4 $ quark hypermultiplets. Such a number of flavors ensures  both the original $SO(7)$ theory being asymptotically free and the $SU(2)$  subgroup being non-asymptotically free. The low-energy gauge group     $U(2) \times U(1) $ is completely broken by the squark VEV's similar to Eq.~(\ref{similar}).
   The large VEV  $\brc \phi_{1} \ckt$    has the form:
 \beq  \brc \phi_{1} \ckt = \left(\begin{array}{ccccccc}0 & i v_0 & 0 & 0 & 0 & 0 & 0 \\-i v_0 & 0 & 0 & 0 & 0 & 0 & 0 \\0 & 0 & 0 & i v_0 & 0 & 0 & 0 \\0 & 0 & -i v_0 & 0 & 0 & 0 & 0 \\0 & 0 & 0 & 0 & 0 & i v_1 & 0 \\0 & 0 & 0 & 0 & -i v_1 & 0 & 0 \\0 & 0 & 0 & 0 & 0 & 0 & 0\end{array}\right), \qquad v_{1} \ne v_{0}.
 \label{SO7vev}\eeqq
 The ``unbroken''  $U(2)$ lies in $SO(4)_{1234}\sim SU(2) \times SU(2)$ while the $U(1)$ factor corresponds to the rotations in the $56$ plane (see Appendix~\ref{a}).
 The semiclassical monopoles   of high-energy  theory
   are \footnote{Within the softly broken ${\cal N}=2$ theory,  the  quantum mechanical vacua with $SU(2) \times  U(1)^{2}$  gauge symmetry, in the limit $m_{i}=m \simeq \Lambda$,    appears to arise from the semiclassical vacua of the form of Eq.~(\ref{SO7vev}), with
  $v_{0}=  m/\sqrt{2} \gg \Lambda,$  $v_{1}=0$, with classical symmetry
   $SU(2) \times  U(1) \times SO(3)_{567}.$
    The $SO(3)_{567}$ gauge sector (pure ${\cal N}=2$ theory) becomes strongly-coupled at low energies  and breaks itself to $U(1)$.   Thus it would be more correct to say $v_{1}\sim \Lambda,$   but then  the discussion about semiclassical monopole  masses $\sim v_{1}/g$, etc., should not be taken too literally.   If one wishes, one could consider a larger gauge group, {\it e.g.},  $SO(9)$, to do a straightforward semiclassical analysis for an unbroken $SU(2)$  group.  In general, the relation between the classical vacua and the fully quantum mechanical vacua is  a rather  subtle issue.  See for instance the discussions in \cite{KMY}.      }
\begin{description}
  \item[(i)]  a triplet of degenerate monopoles of mass $ {2 \,| v_{0}|/g}$  (they arise as in the $SO(5)$ theory discussed above);
  \item[(ii)]  a doublet of degenerate  monopoles of mass ${| v_{0} - v_{1}|}/{g}$: they arise from
  the breaking of $SU_{+}(2) \subset SO(4)_{1256}$ and $SU_{+}(2) \subset SO(4)_{3456}$  (see Appendix~\ref{a});
  \item[(iii)]  a doublet of degenerate   monopoles of mass $ {| v_{0} + v_{1}|}/{g}$: they also arise from
  the breaking of $SU_{-}(2) \subset SO(4)_{1256}$ and $SU_{-}(2) \subset SO(4)_{3456}$;
  \item[(iv)]  a singlet monopole of mass $ {2 \,| v_{1}|/g}$ arising from the breaking of $SO(3)_{567}.$
 \end{description}
Which of these semiclassical monopoles are the lightest and  which of them are stable against decay into lighter monopole pairs,  depend  on the various   VEVs.   It is possible that  the monopoles (ii) or (iii)  are the lightest of all.    Of course  more detailed issues such as which of the degeneracies survives quantum effects,  are questions which go beyond the semiclassical  approximations.

In fact,    when $ v_{0}, v_{1} \sim \Lambda$
the standard semi-classical reasoning fails to give any reliable answer:  a fully quantum-mechanical analysis is needed.      Fortunately, in the softly broken ${\cal N}=2$  theory such analyses have been performed  \cite{CKKM}  and    we do know that the
light   monopoles  in the fundamental representation (${\underline 2}$) of $SU(2)$ appear  in an appropriate vacuum.

Knowing this, we might try to understand how such a result may follow from our definition of the dual group.   At low energies  the gauge group $U(2) \times U(1) $ is completely broken,  leaving a color-flavor diagonal $SU(2)_{C+F}$ symmetry unbroken.
The  theory possesses vortices of
\beq  \pi_{1} (U(2) \times U(1) ) = {\mathbb Z} \times {\mathbb Z}.  \label{homot}
\eeqq
The minimal vortices corresponding to $\pi_{1}(U(2))= {\mathbb Z}$   transform as a  ${\underline 2}$  of  $SU(2)_{C+F}$.

A minimum element of  $\pi_{1} (U(2) \times U(1))$   such as an angle $ 2 \pi$ rotation in the $U(1)_{56}$ factor,  or the minimal  $U(2)$ loop,    corresponds to  vortices  stable in the full theory.  They  would confine   Dirac monopoles associated with $\pi_{1}(SO(7))= {\mathbb Z}_{2}$, if the latter were  introduced in the theory.

The regular monopoles in which we are interested in, are instead  confined by some non-minimal ($k=2$) vortices of the low-energy theory.   However, in contrast to the $SO(5)$  theory discussed in the preceding subsection,  this {\it does not}  necessarily  imply a second-rank tensor representation   of $SU(2)_{C+F}$  of    these  monopoles.
In fact,   the monopoles  of the (ii) group, for instance,  carry the minimum charge of $U(2)$ and  an unit charge of $U(1)$.     Therefore, the relevant $k=2$ vortex corresponds to the minimum  element both of  $\pi_{1} (U(2)) $ and  of $ \pi_{1}(U(1))$,      generated by a $2\pi$ rotation in the $56$ plane together with a minimal loop of $\pi_{1}(U(2))$, analogous to the one discussed in the preceding subsection. As a consequence the monopoles confined by such vortices, by our discussion of Section 3, transform as a {\it doublet}   of the dual group
${\widetilde {SU}}(2) \sim SU(2)_{C+F}$.

 This discussion  naturally generalizes  to all other cases with symmetry breaking,
 $SO(2N+1) \to SU(r) \times U(1)^{N-r+1}  \to \mathbbm {1}$, $r < N$.    The dual magnetic $ SU(r) $  group  observed in the low-energy
 effective theory \cite{CKKM},  under which the light  monopoles transform as a fundamental multiplet,  thus matches
nicely   with the properties of the dual  ${\widetilde {SU}}(r) \sim  SU(r)_{C+F}$  group.

The cases of   $SO(2N) \to SU(r) \times U(1)^{N-r+1}  \to \mathbbm {1}$, $r < N-1$  are similar.  We expect that there is a qualitative difference between the  breaking with the maximum (or next to the maximum)  rank $SU$ factor and  smaller $SU(r)$ unbroken groups.  Such a difference
is indeed observed in the fully quantum mechanical analysis of $SO(N)$  theory \cite{CKKM}.

The behavior of monopoles in  asymptotic-free $USp(2N)$    theories ($N_{f}< 2N+2$)   is more similar to those appearing in  the $SU(N)$ theories,   because of the property, $\pi_{1}(USp(2N))=\mathbbm {1}$.  All monopoles are regular monopoles   due to the partial symmetry breaking,  $USp(2N) \to SU(r) \times U(1)^{N-r+1},\,\,$  $r\le N$.
The transformation property of these monopoles, in the theory with exact unbroken  $SU(r)_{C+F}$  global symmetry,  is deduced from the transformation properties among the non-Abelian vortices  of the low-energy  system $SU(r) \times U(1)^{N-r+1}  \to  \mathbbm {1}:$
they transform as ${\underline r}$  of $SU(r)_{C+F}$.
 Such a result is consistent dynamically, as long as  $r \le    N_{f}/2.$   It is comfortable that these are precisely what is
found from the quantum mechanical analysis \cite{CKM}.

\subsection{Other symmetry breaking patterns and GNOW duality}

Before concluding this section, let us add a few remarks on other symmetry breaking patterns such as  $SO(2N+3) \to   SO(2N+1) \times U(1)$  and $USp(2N+2) \to  USp(2N)\times U(1)$, and the resulting GNOW monopoles.   These cases might be interesting as the GNOW dual groups are different from the original one:  the dual of $SO(2N+1)$ is $USp(2N)$ and vice versa. It is possible to analyze these systems, again setting up  models so that the ``unbroken group'' is completely broken at a much lower mass scales by the set of squark VEVs.  Indeed such a preliminary study has been made in \cite{FK}, where the emergence of the GNOW dual is clearly seen.

However,  the quantum fate of these GNOW dual monopoles is unclear.  More precisely,  
 within the concrete ${\cal N}=2$ models we are working on  where the exact quantum fate of the semiclassical monopoles is known from the analyses made at small $m, \mu$ \cite{CKKM},  we {\it know}  that these GNOW monopoles  do not survive  quantum effects. Only the monopoles carrying the quantum numbers of the $SU(r)$ subgroups discussed in the previous subsection appear.  On the other hand,  there is clearly 
a reason why the GNOW monopoles cannot appear at low energies in these cases: the low-energy effective action would have a wrong global symmetry.   GNOW monopoles are not always relevant quantum mechanically \footnote{Seiberg duals of ${\cal N}=1$  supersymmetric theories with various matter contents, 
provide us with more than enough evidence for it. }. 
These and other peculiar (but consistent) quantum properties of non-Abelian monopoles have been recently discussed in \cite{KMY}.

\section{Conclusion}



In this paper we have examined an idea about the  ``non-Abelian monopoles'',   put forward some time ago \cite{ABEK},
more systematically  and  by using some recent results on the non-Abelian {\it vortices}.
According to this idea, the dual transformation  of non-Abelian monopoles occurring in a system with gauge symmetry breaking
$ G   \,\,\,\longrightarrow    \,\,\, H  $  is to be defined by setting the low-energy  $H$ system in Higgs phase, so that the dual  system  is in confinement  phase.  The transformation  law of the  monopoles follows from that of  monopole-vortex mixed configurations in the system
\[
 G   \,\,\,{\stackrel {v_{1}} {\longrightarrow}} \,\,\, H  \,\,\,{\stackrel {v_{2}} {\longrightarrow}} \,\,\,
 \mathbbm {1},\qquad  (v_{1}\gg v_{2})
\]
under an unbroken, exact color-flavor diagonal symmetry $H_{C+F}\sim {\tilde H}$.
The transformation properties of the  regular monopoles  (\changed{classified by }$\pi_{2}(G/H)$) follow from those among the non-Abelian
vortices (\changed{classified by }$\pi_{1}(H)$), via  the isomorphism  $\pi_{1}(G) \sim  \pi_{1}(H)/\pi_{2}(G/H)$.
Our  results,  obtained in the semiclassical approximation (reliable at $v_{1}\gg v_{2} \gg  \Lambda$)  of
softly-broken ${\cal N}=2$ supersymmetric $SU(N)$ and $SO(N)$  theories,  are -- very non-trivially --  found to be consistent with the fully quantum-mechanical low-energy effective action description (valid at $v_{1}, v_{2} \sim  \Lambda$), available in these theories.

 For   $G= SU(N+1)$, $H= U(N)$, $G_{F}=SU(N_{f})$,  $N_{f}\ge 2\, N$,    this argument proves that
the  monopoles  induced by the $G/H$  breaking  transform as ${\underline N}$   of
${\tilde H}= SU(N)$.     Analogous result holds for $G= SU(N+1)$, $H= U(r)$, $G_{F}=SU(N_{f})$,  $r \le N_{f}/2,$   where  the semi-classical  monopoles transform  as in the fundamental  multiplets (${\underline r}$) (as well as some singlets)   of   $SU(r)$.
These results are  in agreement   with what was found in the fully quantum mechanical treatment of the system \cite{APS,CKM}.

For    $G= SO(2N+1)$,   $H= U(r) \times U(1)^{N-r}$, $G_{F}=SU(N_{f})$   (with $r \le N_{f}/2$,  $r<N$)
we find monopoles which transform  in the fundamental  representation  of the dual  ${\widetilde {SU}}(r)   =  SU(r)_{C+F}$  group. This result is  again  consistent with the fully quantum mechanical  analysis of  ${\cal N}=2$ supersymmetric $SO(N)$  models \cite{CKKM} and in agreement with the universality of certain superconformal theories discovered in this context by Eguchi et. al. \cite{Eguchi}.

   In the case of maximal-rank $SU$ subgroup, such as
  $G=SO(5)$,  $H= U(2)$, there is a qualitative difference both  in  our duality argument and in   the full quantum results.
  For instance the set of monopoles found earlier  by E. Weinberg is shown to belong to  a singlet and a  triplet  representations of the dual $ SU(2)$ group,  but their quantum fate is not known.  In supersymmetric models
a renormalization-group argument suggests (and  the explicit analysis of softly broken ${\cal N}=2$ theory shows) that the triplet does not survive the quantum effects,  as long as the underlying  $SO(5)$ theory is asymptotically free.

For  $G= SO(2N)$,   $H= U(r) \times U(1)^{N-r}$, $G_{F}=SU(N_{f})$   the situation is similar. When $r<N-1$, $r \le N_{f}/2$ we  find  monopoles transforming in the $\underline{r}$ representation of the dual ${\widetilde {SU}}(r)   =  SU(r)_{C+F}$, whereas the maximal and next-to-maximal cases, $r=N,N-1$,   encounter the same renormalization-group constraint   as in $SO(2N+1)$.

Finally for $G= USp(2N)$,   $H= U(r) \times U(1)^{N-r}$, $G_{F}=SU(N_{f})$ the picture is very much like in $SU(N+1)$.  We have monopoles in the fundamental representation of the dual ${\widetilde {SU}}(r)   =  SU(r)_{C+F}$ as long as  $N_f\ge 2\, r$.

   Summarizing, in the context of softly-broken ${\cal N}=2$ supersymmetric gauge theories with $SU$,  $SO$ and $USp$   groups, where
   fully quantum mechanical results are available by  combining the various knowledges such as the Seiberg-Witten curves,
   decoupling theorem,  Nambu-Goldstone theorem,  non-renormalization of Higgs branches, ${\cal N}=1$ ADS  instanton superpotential, vacuum counting,
   universality of conformal theories,  etc.,
       our idea on non-Abelian monopoles is in   agreement with these  known exact results.
       Although such an agreement is comfortable, our arguments, based on the
homotopy-map-stability argument on  almost BPS solitons and on some exact symmetries,  should  be of  more general validity.


%

\section*{Acknowledgement}

K.K.  thanks R.~Auzzi, S.~Bolognesi, J.~Evslin, G.~Paffuti, M.~Strassler  and A.~Vainshtein  for useful discussions, and the organizers of  CAQCD (Continuous Advance of QCD) 2006, Minneapolis (May, 2006),   of the  Benasque workshop ``QCD and Strings''  (July, 2006), and of  SCGT06  Workshop, Nagoya (November, 2006),  for stimulating occasions to discuss some of the ideas exposed here.  M.N.,  K.O. and  M.E. thank E. Weinberg for fruitful discussions and KIAS  for warm hospitality. G.M. and W.V. wish to thank the Theoretical HEP Group of the Tokyo
Institute of Technology and the Theoretical Physics Laboratory of RIKEN for their warm hospitality.
The work of M.~E. and K.~O. is supported by Japan
Society for the Promotion of Science under the Post-doctoral Research Program.
The work of N.~Y. is supported by the Special Postdoctoral Researchers
Program at RIKEN.


\appendix

\section{Monopoles in $SO(N)$ theories} \label{a}

Here are some formulae  useful  for the discussion of  Section \ref{sec:son}.  The minimal $SU(2)$  embeddings ({\it i.e.}, with the smallest Dynkin index,
${\rm Tr}\,  T^{a} T^{b}$)   in $SO(N)$ groups are
obtained   through various $SO(4)\subset SO(N)$  subgroups.  For instance the $SU(2)\times SU(2)  \subset SO(5)$ subgroups are
generated  by
\beq
T_1^{\pm } = -\frac{i }{ 2} ( \Sigma_{23} \pm  \Sigma_{41} ), \quad
T_2^{\pm } = -\frac{i }{ 2} ( \Sigma_{31} \pm  \Sigma_{42} ), \quad
T_3^{\pm } = -\frac{i }{ 2} ( \Sigma_{12} \pm  \Sigma_{43} ),
\eeqq
where  {\it e.g.}
\[ \Sigma_{23} =  \left(\begin{array}{cc}0 & 1 \\-1 & 0\end{array}\right),  \]       is a rotation in the $23$ plane.
 Non-minimal  embeddings correspond  to various $SO(3)$ subgroups, acting in $125$ and $345$ subspaces,  for instance, in the $SO(5)$ example.

The VEV Eq.~(\ref{SOvev}) is proportional to $T_{3}^{+}$: it  leaves  $SU_{-}(2) \times U_{+}(1)$ unbroken.  An $SO(5)$ solution can be obtained  \cite{BS,EW} by embedding the 't Hooft-Polyakov monopoles \cite{TH}   in the  broken $SU(2)$
 as ($S_{a}\equiv  T_{a}^{+}$)
\begin{equation}   A_i({\bf r})  =  A_i^a({\bf r},  {\bf h} \cdot {\bf \alpha}) \, S_a;  \qquad \phi({\bf r}) =
  \chi^a({\bf r},  {\bf h} \cdot {\bf
\alpha})
\, S_a   +  [ \,  {\bf h}   -   ({\bf h}
\cdot {\bf \alpha}) \,    {
\bf
\alpha}^{*}  ]
\cdot {\bf H},
\label{NAmonopol}\end{equation}
where
\begin{equation}    A_i^a({\bf r}) =  \epsilon_{aij}  \frac{ r^j }{ r^2}  A(r); \qquad   \chi^a({\bf r}) =  \frac{ r^a }{ r} \chi(r), \qquad    \chi(\infty)=
  {\bf h} \cdot {\bf \alpha}.
\end{equation}
    Note that $\phi({\bf r}=(0,0,\infty) ) = \phi_0.$
In the above formula the   Higgs field vacuum expectation value  (VEV)    has been   parametrized in  the form
\begin{equation}     \phi_0  =  {\bf h} \cdot  {\bf H},
\end{equation}
where  $ {\bf h} = (h_1, \ldots, h_{{\small{\textup{rank}(G)}}})$
 is a constant vector representing
the VEV.  The root vectors  orthogonal to $ {\bf h}$   ($\propto \alpha $ in Fig. \ref{roots})   belong to the unbroken  subgroup   $H$   ($\gamma$  in Fig. \ref{roots}).

The above consideration is basically group-theoretic and is valid in any types of theories, supersymmetric or not. Now we specialize to the concrete dynamical models we are working on:   ${\cal N}=2$ supersymmetric gauge theories.
Under the symmetry breaking  $SO(5)\to U(2)$  the quark superfields $Q$ and ${\tilde Q}$  in the first four components of the vector representation  rearrange themselves as follows.  Recall  that  the  relevant superpotential terms have the form,  $Q ( m {\mathbf 1} + \sqrt{2} \Phi ) {\tilde Q} $,  summed over  diagonal flavor indices, $A=1,2,\ldots, N_{f}$ left implicit.   For each flavor,  the adjoint scalar VEV of the form Eq.~(\ref{SOvev}), with   $v= m/\sqrt{2}$,  gives rise to  a  $2 \times 2$  block-diagonal   mass matrix
\beq  m {\mathbf 1} + \sqrt{2} \Phi  =\left(\begin{array}{cc}{\bf  V} & 0 \\0 & {\bf V}\end{array}\right)\qquad   {\bf V}  =m\,  \left(\begin{array}{cc}1 & i \\-i & 1\end{array}\right)
\eeqq
in  color.   ${\bf V}$   has  one vanishing and one massive eigenvalues.   Thus the four  fields
\beq {\hat Q}^{1} =  \frac{1}{\sqrt{2}} ( Q^{1} + i Q^{2}), \quad  {\hat Q}^{3} =  \frac{1}{\sqrt{2}} ( Q^{3} + i Q^{4}),  \quad
{\hat {\tilde Q}}^{1} =  \frac{1}{\sqrt{2}} ( {\tilde Q}^{1} + i {\tilde Q}^{2}), \quad  {\hat {\tilde Q}}^{3} =  \frac{1}{\sqrt{2}} ( {\tilde Q}^{3} + i {\tilde Q}^{4}),  \non \eeqq
are massless.  The orthogonal combinations such as   $ \frac{1}{\sqrt{2}} ( Q^{1} -  i Q^{2}) $ become massive and decouple from the low-energy theory.

The massless quark  superfields of the low-energy  $SU(2)$ theory are the combinations
\beq   q^{1} =   \frac{1}{\sqrt{2}}  (    {\hat Q}^{1}  + i  {\hat Q}^{3} );   \qquad   q^{2} =   \frac{1}{\sqrt{2}}  ( i {\hat Q}^{1} +  {\hat Q}^{3} ),
\eeq
which form a  ${\underline 2}$,   and
\beq   {\tilde q}^{1} =   \frac{1}{\sqrt{2}}  (    {\hat {\tilde Q}}^{1}  -   i  {\hat  {\tilde Q}}^{3} );   \qquad  {\tilde q}^{2} =   \frac{1}{\sqrt{2}}  (-   i {\hat  {\tilde Q}}^{1} +  {\hat  {\tilde Q}}^{3} ),
\eeq
which form a  ${\underline 2^{*}}$. \footnote{ For a general change of basis vectors  from $SO(2N)$ to  $U(N)$ see the Appendix A of \cite{CKM}.   }

It is straightforward to generalize
the above construction to    $SO(2N+1) \to SU(r) \times U(1)^{N-r+1}$,  $r < N$.
$N_{f}$   quark hypermultiplets in the $SO(2N+1)$ vector representation yield  precisely $N_{f}$ flavors of massless quarks
in ${\underline r}$ of  $SU(r) $  plus a number of singlets.

\newcommand{\J}[4]{{\sl #1} {\bf #2} (#3) #4}
\newcommand{\andJ}[3]{{\bf #1} (#2) #3}
\newcommand{\AP}{Ann.\ Phys.\ (N.Y.)}
\newcommand{\MPL}{Mod.\ Phys.\ Lett.}
\newcommand{\NP}{Nucl.\ Phys.}
\newcommand{\PL}{Phys.\ Lett.}
\newcommand{\PR}{ Phys.\ Rev.}
\newcommand{\PRL}{Phys.\ Rev.\ Lett.}
\newcommand{\PTP}{Prog.\ Theor.\ Phys.}
\newcommand{\hep}[1]{{\tt hep-th/{#1}}}

\end{document}